\documentclass{ifacconf}

\usepackage{natbib}
\usepackage[utf8]{inputenc}     

\usepackage[T1]{fontenc}        
\usepackage{lmodern}
\usepackage[english]{babel}              

\usepackage{graphicx}           
\usepackage{amsmath,amssymb}    
\allowdisplaybreaks

\usepackage{color}
\usepackage[usenames]{xcolor}   
\usepackage{microtype} 

\usepackage{marginnote}         

\usepackage{mathtools}
\usepackage{siunitx}
\usepackage{pbox}
\usepackage{algorithm}

\usepackage{tikz}				
\usetikzlibrary{arrows}				
\usetikzlibrary{shapes.misc}
\usetikzlibrary{patterns}
\usepackage{pgfplots}
\usetikzlibrary{calc}

\colorlet{istorange}{orange}
\colorlet{istgreen}{green!50!black}
\colorlet{istblue}{blue} 
\colorlet{istred}{red!90!black}

\newtheorem{assumption}{Assumption}

\newtheorem{remark}{Remark}
\newtheorem{Algorithm}{\bf Algorithm}

\newcommand{\X}{\mathbb{X}}

\newcommand{\U}{\mathbb{U}}

\newcommand{\I}{\mathbb{I}}
\newcommand{\R}{\mathbb{R}}

\renewcommand{\qed}{\hfill\ensuremath{\square}}
\begin{document}
\begin{frontmatter}
	

\title{A Resource-Aware Approach to Self-Triggered Model Predictive Control: Extended Version} 

\thanks[footnoteinfo]{This work is an extended version of \cite{Wildhagen20_2}. Funded by the Deutsche Forschungsgemeinschaft (DFG, German Research Foundation) - 285825138; 390740016 and the Swiss National Science Foundation under the RISK project (grant 200021 175627).}

\author[First]{Stefan Wildhagen} 
\author[Second]{Colin N. Jones} 
\author[First]{Frank Allg{\"o}wer}

\address[First]{Institute for Systems Theory and Automatic Control, 
   University of Stuttgart, Germany (\{wildhagen,allgower\}@ist.uni-stuttgart.de).}
\address[Second]{Automatic Control Laboratory, 
   Ecole Polytechnique Federale de Lausanne (EPFL), Switzerland (colin.jones@epfl.ch).}

\begin{abstract}                
: In this paper, we consider a self-triggered formulation of model predictive control. In this variant, the controller decides at the current sampling instant itself when the next sample should be taken and the optimization problem be solved anew. We incorporate a pointwise-in-time resource constraint into the optimization problem, whose exact form can be chosen by the user. Thereby, the proposed scheme is made resource-aware with respect to a universal resource, which may pertain in practice for instance to communication, computation, energy or financial resources. We show that by virtue of the pointwise-in-time constraints, also a transient and an asymptotic average constraint on the resource usage are guaranteed. Furthermore, we derive conditions on the resource under which the proposed scheme achieves recursive feasibility and convergence. Finally, we demonstrate our theoretical results in a numerical example.
\end{abstract}


\end{frontmatter}

\section{Introduction}

Model predictive control (MPC) is an effective method for the control of nonlinear, constrained and multi-variable systems, and has received great attention in both theory (\cite{Mayne00}) and practice (\cite{Qin03}). The main principle of MPC is receding horizon optimal control: an optimal control problem is solved over a finite horizon and the first part of the predicted input is applied to the process, before a new sample of the state is taken and the scheme is repeated. Traditionally, these samples are taken equidistantly in time although more recently, some approaches abandoned this concept in favor of aperiodic sampling. These works considered mostly MPC in the context of Networked Control Systems (NCS), where state measurements and control signals must be transmitted via communication networks with a limited amount of communication resources. In such an environment, to avoid network congestion, samples should be taken and new control inputs should be transmitted only when needed.

To determine when samples should be taken in aperiodic MPC for NCS, mainly event-triggered and self-triggered approaches are of interest. Event-triggered MPC (see for example \cite{Varutti09,Eqtami10,Li14}) uses a triggering condition, which is persistently monitored during runtime, in order to determine when the next sample should be taken. In contrast, in self-triggered MPC, the next sampling time is determined already at the current sampling time based on predictions of the system behavior. Some of the earliest approaches in this direction are presented in \cite{Bernardini12,Berglind12} and \cite{Henriksson12}. In the former two works, the sampling period is chosen as long as possible such that in closed loop, stability and a certain level of performance are still guaranteed. In the latter work on the other hand, the next sampling instant is determined based on an optimization problem, which jointly considers in its objective function the process cost and the communication load. In such an approach, as taken also in \cite{Henriksson15,Zhan17,Li17,Li18}, the controller may directly trade off between control performance and length of the sampling interval.

In the latter works, the input is additionally held constant in between sampling instances. In effect, these self-triggered MPC schemes trigger transmissions of state measurements and new inputs over the communication network only when required by the process, thereby being resource-aware with respect to the network's communication resources. In practical applications, however, other types of resources are of interest as well: in battery-operated devices, energy is scarce; in cloud computing, there is a financial component to how much computational power is requested; on micro controllers, computing performance is considerably limited.

In this work, we propose a self-triggered MPC which determines the next sampling time via optimization similar to the above-mentioned schemes, although only the process cost is represented in its objective function. Hence, in contrast to these works (which add a cost term on the sampling interval), the proposed controller chooses the sampling interval that is most favorable for the process itself. To achieve resource awareness nonetheless, a \textit{constraint} on the resource is added to the optimization problem. In particular, we consider a resource that is charged at a constant rate and that, using a certain sampling interval, is decreased by a corresponding cost. Surplus resources are saved in a storage to be used at a later point in time. The resource constraint is then to enforce that the level of the resource does not fall below zero. Since we allow for arbitrary resource cost functions, the proposed self-triggered MPC is resource-aware with respect to a general resource. The concrete form of the resource cost function may be chosen by the user, such that energy-related, financial, computational or other types of resources can be incorporated into the controller.

A self-triggered MPC may adaptively change its sampling interval based on the current operating conditions of the process, and thereby react to unforeseen operating conditions such as a set point change or disturbances. Hence, it is in general not possible to bound the average sampling frequency over a certain time interval a priori (apart from trivial bounds by a lower bounded sampling interval). Our main contribution is to show that in the proposed formulation, closed-loop guarantees on the average resource usage are indeed achieved even in unforeseen operating conditions. Thereby, both transient (which hold over arbitrary, but finite time horizons) and asymptotic average constraints on the resource usage are enforced. Especially non-trivial transient bounds are novel in self-triggered MPC. Furthermore, we provide conditions on the process and the resource under which the proposed self-triggered MPC is recursively feasible and achieves convergence.

The remainder of the paper is organized as follows. In Section \ref{sec_preliminaries}, the considered process and resource dynamics are introduced before the self-triggered MPC scheme is presented in Section \ref{sec_control_setup}. The main results on average resource usage, and on recursive feasibility and convergence are derived in Sections \ref{sec_average_constraints} and \ref{sec_guarantees}. A numerical example is given in Section \ref{sec_num_ex} before the work is concluded in Section \ref{sec_summary}.

Apart from standard notation, let $\I$ denote the set of integers and for $a,b\in\I$, let us define $\I_{[a,b]}\coloneqq\I\cap[a,b]$ and $\I_{\ge a}\coloneqq\I\cap[a,\infty)$.

\section{Preliminaries} \label{sec_preliminaries}

\subsection{System description}

We consider a nonlinear continuous-time process
\begin{equation}
\dot{x}(t) = f(x(t),u(t)), \; x(0)\in\X_0, \label{plant}
\end{equation}
with time $t\in[0,\infty)$, state $x(t)\in\R^n$, and input subject to the constraints $u(t)\in\U\subseteq\R^m$. We have for the vector field $f:\R^n\times\U\rightarrow\R^n$, and furthermore, there is a cost $\ell:\R^n\times\U\rightarrow [0,\infty)$ associated with the process. The following standing assumptions are borrowed from \cite{Findeisen06} and \cite{Faulwasser12}, and will be assumed to hold true for the remainder of the paper without further mention.
\begin{assumption}
	The input constraint set $\U$, the vector field $f$ and the cost $\ell$ fulfill the following properties:
	\begin{itemize}
		\item $\U$ is compact and contains the origin.
		\item $f$ is continuous and locally Lipschitz in $x$ for any $u\in\U$, and fulfills $f(0,0)=0$.
		\item For any $x_0\in\X_0$ and any input function $u(\cdot)\in\mathcal{PC}(\U)$ (the set of piecewise continuous input functions with values in $\U$), \eqref{plant} has an absolutely continuous solution.
		\item $\ell$ is continuous and positive definite.
	\end{itemize}
\label{ass_standing}
\end{assumption}

\subsection{Self-triggering mechanism and ZOH Input}

The MPC is activated at discrete sampling instants $t_k, k\in \I_{\ge 0}$, which are not evenly spaced in time as in standard formulations. Instead, the MPC decides about the  next sampling instant itself in a self-triggered fashion. The sampling interval or inter-sampling period $\Delta_k\coloneqq t_{k+1}-t_k\ge 0$ is hence a part of the optimization variables.

In between sampling times, the control law $\pi(x)$ is implemented in a zero-order hold fashion with constant input values during the sampling interval, i.e.,
\begin{equation*}
u(t)=\pi(x(t_k)), \; t\in[t_k,t_{k+1}).
\end{equation*}
As a result, the considered control algorithm achieves a sparsely changing control input.

\subsection{Resource constraints}

In this work, we assume that there is a limited resource which determines when the controller may be activated and the control input may be changed. The level of the resource $r$ is charged at a constant rate of $p\in[0,\infty)$ and using a certain sampling interval incurs a corresponding resource cost of $\mu(\Delta)$, $\mu:[0,\infty)\rightarrow[0,\infty)$. The resource level does not exceed a certain upper threshold $\overline{r}\in[0,\infty)$. Hence, the resource level at sampling times evolves according to
\begin{equation}
r_{k+1} = \min\{r_k + \Delta_k p - \mu(\Delta_k),\overline{r}\}, \; r_0 \in [0,\overline{r}]. \label{sd_resource}
\end{equation}
A certain $\Delta_k$ is only feasible if the resource is not depleted using this sampling interval, i.e., the resource level is constrained by
\begin{equation}
r_k \ge 0, \; \forall k\in\I_{\ge 0}. \label{resource_constraint}
\end{equation}

Such a form of resource dynamics is inspired by the token bucket specification, which is a common model in communication theory for the communication resources of a digital network (see \cite{Tanenbaum11}). Recently, this model was used in the context of NCS to determine when new inputs should be transmitted over a resource-constrained network (\cite{Linsenmayer18,Wildhagen19_2}). In the token bucket specification, the resource cost $\mu$ is assumed to be independent of the sampling interval. By the means of an arbitrary resource cost as used in this work, more general resource-constrained setups than NCS can be considered.

Importantly, if the resource evolving according to \eqref{sd_resource} fulfills constraint \eqref{resource_constraint}, then one can also guarantee constraints on the average resource usage. This is useful, e.g., if one wants to prescribe a certain average sampling frequency or energy consumption. We will detail this idea in Section \ref{sec_average_constraints}.

\section{Self-triggered MPC Scheme} \label{sec_control_setup}

In this section, we present the operation scheme of the proposed self-triggered MPC and discuss some of its properties. The MPC optimization problem introduced here provides no guarantees on recursive feasibility and convergence for the resulting closed-loop system. We will present an appropriate extension in order to provide theoretical guarantees on these matters in Section \ref{sec_guarantees}.

The main idea is to use a sampled-data version of the process \eqref{plant} in the MPC optimization problem. Since the input is assumed to be constant in between two sampling instants, the state of the system after one sampling interval is determined only by the initial condition, the input and the inter-sampling period according to
\begin{equation}
x_{k+1}=\phi(x_k,u_k,\Delta_k)\coloneqq \int_{0}^{\Delta_k} \hspace{-5pt} f(z(t),u_k) \mathrm{d}t, \; x_0=x(0)
\label{sd_plant}
\end{equation}
where
\begin{equation}
\dot{z}(t)=f(z(t),u_k), \; z(0)=x_k.
\label{z_t}
\end{equation}
This sampled-data state naturally coincides with the state of the continuous-time process \eqref{plant} at the sampling instances. Similarly, a sampled-data version of the cost can be defined
\begin{equation}
\lambda(x_k,u_k,\Delta_k) \coloneqq \int_{0}^{\Delta_k} \hspace{-5pt} \ell(z(t),u_k) \mathrm{d}t.
\label{sd_cost}
\end{equation}

The MPC optimization problem then minimizes the cost \eqref{sd_cost} along the predicted trajectories of the sampled-data process \eqref{sd_plant}. Let $\vec{x}(t_k)\coloneqq\{x_0(t_k),\ldots,x_{N}(t_k)\}$ and $\vec{r}(t_k)\coloneqq\{r_0(t_k),\ldots,r_{N}(t_k)\}$ denote the predicted states and resource levels over the prediction horizon $N\in\I_{\ge 1}$. Furthermore, let $\vec{u}(t_k)\coloneqq\{u_0(t_k),\ldots,u_{N-1}(t_k)\}$ and $\vec{\Delta}(t_k)\coloneqq\{\Delta_0(t_k),\ldots,\Delta_{N-1}(t_k)\}$ denote the predicted inputs and sampling intervals. Then, at sampling time $t_k$, the controller solves the MPC optimization problem
\begin{subequations}
\begin{align}
&\min_{\substack{\vec{x}(t_k),\vec{r}(t_k) \\ \vec{u}(t_k),\vec{\Delta}(t_k)}}  \underbrace{\sum_{i=0}^{N-1} \lambda(x_i(t_k),u_i(t_k),\Delta_i(t_k))}_{\eqqcolon V(\vec{x}(t_k),\vec{r}(t_k),\vec{u}(t_k),\vec{\Delta}(t_k))} \nonumber \\
\text{s.t. }	&x_{i+1}(t_k)=\phi(x_{i}(t_k),u_{i}(t_k),\Delta_i(t_k)) \label{opt_const_dyn_plant} \\
&r_{i+1}(t_k) \hspace{-1pt}=\hspace{-1pt} \min\{ r_i(t_k) + p\Delta_i(t_k) - \mu(\Delta_i(t_k)),\overline{r} \} \label{opt_const_dyn_resource} \\
&r_i(t_k) \ge 0, \; u_i(t_k)\in\U \label{opt_const_constraints}  \\
&\underline{\Delta}\le\Delta_i(t_k)\le\overline{\Delta}, \quad \forall i\in\I_{[0,N-1]} \label{opt_const_interval_normalization} \\
&x_0(t_k)=x(t_k), \; r_0(t_k)=r_k. \label{opt_const_IC}
\end{align}
\end{subequations}
In this problem, the independent optimization variables are the inputs $u_i(t_k)$ as well as the sampling intervals $\Delta_i(t_k)$. The predicted state and resource level trajectories must fulfill their respective dynamics and initial conditions as enforced in \eqref{opt_const_dyn_plant}, \eqref{opt_const_dyn_resource} and \eqref{opt_const_IC}. Additionally, the predicted resource levels and inputs must comply with their respective constraints \eqref{opt_const_constraints}. Lastly, a lower and upper bound $\underline{\Delta},\overline{\Delta}\in[0,\infty)$ on the sampling intervals are imposed \eqref{opt_const_interval_normalization}. 
\begin{remark}
Note that the state transition map $\phi$ does not have to be determined analytically in order to solve the MPC optimization problem. In practice, one may rely on numerical integration to determine $\phi$ approximately.
\end{remark}
\begin{remark}
	The lower and upper limits on the inter-sampling period are not necessary to guarantee recursive feasibility later. In contrast, they need to be carefully adjusted not to jeopardize this property, as we will see in Section \ref{sec_guarantees}. Then again, the upper limit guarantees a minimal attention devoted to the process and may facilitate finding a solution to the optimization problem in practice. The lower limit may be used to avoid Zeno behavior.
\end{remark}
\begin{remark}
	Since the MPC optimization problem relies entirely on a sampled-data formulation, state constraints on the continuously evolving process state cannot be incorporated. At sampling times, state constraints of the form $x(t)\in\X\subseteq\R^n$ may be enforced according to
	\begin{equation*}
	x_i(t_k)\in\X, \quad \forall i\in\I_{[0,N-1]}.
	\end{equation*}
	Although in theory, a guarantee on recursive feasibility would not be lost under the assumptions presented in Section \ref{sec_guarantees}, such a constraint provides no guarantee that the closed-loop state trajectory evolving in continuous time does not violate the state constraints.
	\label{rem_state_constraints}
\end{remark}
\begin{remark}
	In the MPC optimization problem, the sampling intervals may take values in a connected set. In contrast, previous works were formulated in discrete time, such that the sampling intervals took values in a discrete set giving rise to a mixed integer optimization problem.
\end{remark}

In self-triggered MPC, the elapsed time within the prediction horizon is subject to the chosen sampling intervals instead of being predetermined. Previous formulations of self-triggered MPC have always considered one sampling instant as an independent optimization variable within the prediction horizon, i.e., $N=1$ (cf. \cite{Henriksson12,Henriksson15,Zhan17,Li17,Li18}). The proposed approach generalizes this concept and considers an arbitrary but fixed number of sampling instants $N\ge 1$. It should be noted that using a general $N\ge 1$ was hinted at in \cite[Remark 7]{Henriksson15}.

The state, resource level, input and inter-sampling period sequences solving the MPC optimization problem are denoted by the superscript $^*$. The MPC then operates with the following scheme.
\begin{algorithm}
	\begin{Algorithm}\label{scheme_MPC}
		\normalfont{\textbf{MPC Scheme}}
		\begin{enumerate}
			\item[0)] Set $t_0=0$ and $k=0$.
			\item[1)] At time $t_k$, measure $x(t_k)$ and solve the MPC optimization problem.
			\item[2)] Apply $u(t)=u_0^*(t_k), \; \forall t\in[t_k,t_k + \Delta_0^*(t_k))$ to \eqref{plant} and calculate $r_{k+1}=\min \{r_k + p \Delta_0^*(t_k) - \mu(\Delta_0^*(t_k)),\overline{r}\}$.
			\item[3)] Set $t_k\leftarrow t_k + \Delta_0^*(t_k)$, $k\leftarrow k+1$ and go to 1).
		\end{enumerate}
	\end{Algorithm}
\end{algorithm}

\begin{remark}
	As an extension, one could imagine a ``multi-step'' version of Algorithm \ref{scheme_MPC}, where the first $M\in\I_{[1,N]}$ parts of $\vec{u}^*(t_k)$ and $\vec{\Delta}^*(t_k)$ are applied in closed loop. Then the state would be sampled and the MPC optimization problem be solved again at $t_k+\sum_{i=0}^{M-1} \Delta_i^*(t_k)$. Although not further mentioned, all following results also hold under application of such a scheme.
\end{remark}

\section{Average Constraints on Resource Usage} \label{sec_average_constraints}

Under application of Algorithm \ref{scheme_MPC}, apparently the closed-loop inter-sampling periods $\Delta_k\coloneqq\Delta_0^*(t_k)$ are such that the resource constraint \eqref{resource_constraint} is always fulfilled. With this observation, an upper limit on the average resource usage under application of the MPC can be established.
\begin{lem}
	Suppose that for a $k\in\I_{\ge 1}$, the MPC optimization problem is feasible at all sampling instances $t_i$, $i\in\I_{[0,k-1]}$. Then,
	\begin{equation}
	\tfrac{\sum_{i=0}^{k-1}\mu(\Delta_i)}{t_k} \le \tfrac{r_0}{t_k} + p \label{no_zeno}
	\end{equation}
	holds for the closed loop under application of Algorithm \ref{scheme_MPC}.
	\label{lem_average_constraint_finite}
\end{lem} \vspace{-10pt}
\begin{pf}
	From the resource dynamics \eqref{sd_resource}, we obtain
	\begin{align*}
	r_k&=\min\{r_{k-1} + \Delta_{k-1} p - \mu(\Delta_{k-1}),\overline{r}\} \\
	&\le r_{k-1} + \Delta_{k-1} p - \mu(\Delta_{k-1}).
	\end{align*}
	Repeating this argument and considering \eqref{resource_constraint} gives
	\begin{equation*}
	0\le r_k \le r_0 + \textstyle\sum_{i=0}^{k-1} \Delta_i p - \mu(\Delta_i).
	\end{equation*}
	Because $\sum_{i=0}^{k-1} \Delta_i = t_k$, we obtain \eqref{no_zeno} by rearranging. \qed
\end{pf}
\begin{remark}
	Note that Lemma \ref{lem_average_constraint_finite} also implies that $$\tfrac{\sum_{i=j}^{k-1}\mu(\Delta_i)}{t_k-t_j} \le \tfrac{r_j}{t_k-t_j} + p$$ for any $j\in\I_{[0,k-1]}$. In this respect, \eqref{no_zeno} can be seen as a transient average constraint, i.e., a constraint on a quantity averaged over any finite time period (\cite{Mueller14b}). 
\end{remark}
With the following assumption, one can also bound the asymptotic average resource usage.
\begin{assumption}
	Either $\underline{\Delta}=0$ and $\mu(0)>0$ or $\underline{\Delta}>0$ holds.

	\label{ass_no_zeno}
\end{assumption}
\begin{thm}
	Consider Lemma \ref{lem_average_constraint_finite} and in addition, suppose that Assumption \ref{ass_no_zeno} holds. Then, $\lim_{k\rightarrow\infty} t_k = \infty$ and
	\begin{equation}
	\lim_{k\rightarrow\infty} \tfrac{\sum_{i=0}^{k-1}\mu(\Delta_k)}{t_k} \le p. \label{average_constraint}
	\end{equation}
	\label{lem_average_constraint}
\end{thm} \vspace{-13pt}
\begin{pf}
	First, assume that $\underline{\Delta}=0$ and $\mu(0)>0$. Starting from a finite initial resource level $r_0$, the controller may pick $\Delta^*(t_k)=0$ only at finitely many consecutive sampling instances. This is due to the fact that the resource depletes as it has no time to fill up again and $\mu(0)>0$. After that, some $\Delta^*(t_k)>0$ must be used. Repeating this argument gives $\lim_{k\rightarrow\infty} t_k = \infty$. Second, consider the case $\underline{\Delta}>0$, for which establishing $\lim_{k\rightarrow\infty} t_k = \infty$ is trivial.
	
	Since $r_0$ is finite, the claim follows readily from \eqref{no_zeno}. \qed
\end{pf}
Lemma \ref{lem_average_constraint_finite} and Theorem \ref{lem_average_constraint} establish that, by enforcing the resource constraint \eqref{resource_constraint} at each point in time, the proposed self-triggered MPC also achieves the transient average constraint \eqref{no_zeno} and the asymptotic average constraint \eqref{average_constraint} on the resource usage. Note that \eqref{no_zeno} and \eqref{average_constraint} hold for any scheme that guarantees the resource constraint \eqref{resource_constraint} under the dynamics \eqref{sd_resource} in closed loop, independent of unforeseen operating conditions. MPC schemes which considered resource usage were formerly discussed in the classical setup in \cite{Gommans17} and in the self-triggered setup in \cite{Henriksson12,Henriksson15,Zhan17,Li17,Li18}. In the former work, a periodically sampled MPC was considered, optimizing the transmission schedule of new inputs over a resource-constrained communication network. An asymptotic average constraint on the transmission frequency $\lim_{k\rightarrow\infty}\frac{k}{t_k} \le q$ was guaranteed, simply by imposing the same average resource usage over each sampling period. In the latter works, the trivial upper bound on the sampling rate $\frac{k}{t_k} \le \frac{1}{\underline{\Delta}}$ (which holds for the proposed scheme as well) holds by design. In addition, these schemes achieve in the limit $\lim_{k\rightarrow\infty}\frac{k}{t_k} = \frac{1}{\overline{\Delta}}$. 

The approach proposed in this paper generalizes the incorporation of average resource constraints into resource-aware MPC in two directions: First, the proposed approach provides a higher flexibility in the scheduling of sampling instances than previous approaches. In view of \eqref{no_zeno}, an excess of resource usage over \textit{any} finite time horizon (as long as \eqref{resource_constraint} is fulfilled) can be compensated for in finite time with a later phase of little resource usage. In \cite{Gommans17}, the resource usage can only be redistributed within the considered sampling period, while in the self-triggered setups, an excess of resource usage is only compensated for in infinite time. A non-trivial transient average constraint as given by \eqref{no_zeno} is even novel in self-triggered MPC. Second, since an arbitrary resource cost $\mu$ is considered, more general constraints can be enforced by virtue of \eqref{no_zeno} and \eqref{average_constraint} than only bounding the sampling rate $\frac{k}{t_k}$. Here, the physical interpretation of the resource constraints depends on the cost function, which may be given by the setup or can be chosen by the user. Three examples for the interpretation of the average resource constraints \eqref{no_zeno} and \eqref{average_constraint} in a technical context are given below.

\textbf{Example 1.} Consider that updates of state measurements and inputs need to be transmitted between sensors and the controller, respectively the controller and actuators via a communication network fulfilling the token bucket specification. In this case the resource cost is $\mu(\Delta)=c$, $c\in[0,\infty)$, such that an upper bound on the average sampling frequency is enforced: $\frac{k}{t_k}\le \frac{r_0}{t_k c}+\frac{p}{c}$. Due to the fact that the input changes only at sampling times and the self-triggering setup, the same bound holds for the average transmission frequency of control inputs and state measurements over the communication network.

\textbf{Example 2.} Consider that there is an energy resource, where the resource cost $\mu$ is related to the energy that is consumed by a processing unit to solve the MPC optimization problem within a certain sampling interval. Typically in processing units, the energy $\mu$ required for computation is inversely proportional to the computation time $\Delta$. Since the MPC may decide on its computation time in the self-triggered setup, one may place a constraint on the average power consumption $\frac{\sum \mu}{t}$ of the control: $\lim_{k\rightarrow\infty}\frac{\sum \mu(\Delta_k)}{t_k}\le p$. Constraining the asymptotic power consumption is especially meaningful in battery-operated devices, e.g., drones, where the control unit may account for a substantial part of the overall energy usage.

\textbf{Example 3.} Large-scale MPC problems requiring extensive computation and data resources can be offloaded onto cloud computing systems. Utilizing distributed computing approaches (e.g., ADMM, AMA, etc), the time required to solve an MPC optimization problem and thereby compute the control input is inversely proportional to the number of processors utilized. A strength of cloud computing is the ability to rapidly scale up or down the number of processors utilized, with a cost proportional to the number of processors used. In this case, the resource being bounded is the financial cost, and the scheme provides an ability to set a monetary budget on the cost of computation.

\begin{remark}
	Similar concepts of ensuring average constraints by pointwise-in-time constraints are also known in the literature on economic MPC (\cite{Angeli12,Mueller14b}). These approaches use a classical periodically activated MPC to control a discrete-time system.
\end{remark}

\section{Ensuring Recursive Feasibility and Convergence} \label{sec_guarantees}

A well-established way to theoretically guarantee recursive feasibility and stability in MPC is to add a terminal equality constraint to the MPC optimization problem. Although simple, this approach is in general restrictive and leads to a smaller feasible set compared to more advanced methods, e.g., the terminal set method. Nonetheless, we use this formulation here in a first step to provide valuable insights into the requirements that the resource must fulfill in order to guarantee recursive feasibility and convergence.

For the terminal equality constraint setup, the MPC opti- mization problem is denoted by $\mathcal{P}(x(t_k),r_k)$ and reads
\begin{align}
V^*(x(t_k),r_k) &\coloneqq \min_{\substack{\vec{x}(t_k),\vec{r}(t_k) \\ \vec{u}(t_k),\vec{\Delta}(t_k)}}  V(\vec{x}(t_k),\vec{r}(t_k),\vec{u}(t_k),\vec{\Delta}(t_k)) \nonumber \\
\text{s.t. } &\eqref{opt_const_dyn_plant}-\eqref{opt_const_IC} \nonumber \\
&x_N(t_k) = 0, \; r_N(t_k) \ge 0. \label{opt_tec}
\end{align}
Note that constraints on the terminal process state and resource level \eqref{opt_tec} were added, as compared to the MPC optimization problem defined in Section \ref{sec_control_setup}. We denote by $\mathcal{XR}$ the set of all process states and resource levels such that $\mathcal{P}(x,r)$ is feasible.

Next, we analyze the analytic properties of the proposed control scheme. To this end, let us first define $\mathbb{D}$ as the set of sampling intervals which fulfill
\begin{equation}
\mathbb{D}\coloneqq\{\Delta\in[0,\infty)| p\Delta - \mu(\Delta)\ge 0\}. \label{delta_tilde}
\end{equation}
\begin{assumption}
	The set $\mathbb{D}$ is nonempty.
	\label{ass_delta_tilde}
\end{assumption} \vspace{-7pt}
In other words, there exists at least one sampling interval such that the resource fills up to a level where sampling and changing the input is possible.
\begin{assumption}
	The minimum and maximum sampling intervals are such that there exists a $\Delta\in\mathbb{D}$ which fulfills
	\begin{equation*}
	\underline{\Delta} \le \Delta \le \overline{\Delta}.
	\end{equation*}
	\label{ass_min_max_choice}
\end{assumption} \vspace{-15pt}
\begin{thm}
	Suppose that $(x(0),r_0)\in\mathcal{XR}$ and that Assumptions \ref{ass_delta_tilde} and \ref{ass_min_max_choice} hold. Then $\mathcal{P}(x(t_k),r_k)$ under application of Algorithm \ref{scheme_MPC} is feasible for all $k\in\I_{\ge 0}$.
	\label{thm_tec_rec_feas}
\end{thm} \vspace{-6pt}
\begin{pf}
	Assume that $\mathcal{P}$ was feasible at time $t_k$. Denote $\tilde{\Delta}\in\mathbb{D}$ a sampling interval that fulfills Assumption \ref{ass_min_max_choice}. Then at the next sampling instant $t_{k+1}=t_k+\Delta^*_0(t_k)$, consider the candidate input and inter-sampling period sequences
	\begin{equation}
	\begin{aligned}
	\tilde{\vec{u}}(t_{k+1}) &= \{u_1^*(t_k),\ldots,u_{N-1}^*(t_k),0\} \\
	\tilde{\vec{\Delta}}(t_{k+1}) &= \{\Delta_1^*(t_k),\ldots,\Delta_{N-1}^*(t_k),\tilde{\Delta}\},
	\end{aligned}
	\label{feas_inp}
	\end{equation}
	which result in the state and resource level sequences
	\begin{align}
	\tilde{\vec{x}}(t_{k+1}) &= \{x_1^*(t_k),\ldots,x_N^*(t_k),\phi(x_N^*(t_k),0,\tilde{\Delta})\} \nonumber \\
	\tilde{\vec{r}}(t_{k+1}) &= \{r_1^*(t_k),\ldots,r_{N}^*(t_k), \label{feas_state} \\
	&\hspace{40pt}\min\{r_{N}^*(t_k)+p\tilde{\Delta}-\mu(\tilde{\Delta}),\overline{r}\}\}. \nonumber
	\end{align}
	Due to $x_N^*(t_k)=0$ and $f(0,0)=0$, we have $\phi(x_N^*(t_k),0,\tilde{\Delta})$\\$=x_N^*(t_k)=0$. Furthermore, $r_{N}^*(t_k)+p\tilde{\Delta} - \mu(\tilde{\Delta}) \ge r_{N}^*(t_k)\ge 0$ according to \eqref{delta_tilde}, such that \eqref{opt_tec} is fulfilled by these sequences. The remaining constraints are also fulfilled since $0\in \U$ and $\underline{\Delta} \le \tilde{\Delta} \le \overline{\Delta}$. \qed
\end{pf}

Before we comment on convergence, we first state the following intermediate result on the sampled-data cost.
\begin{lem}
	It holds that $\lambda(0,0,\Delta)=0$ for all $\Delta\in[0,\infty)$.
	\label{lem_sd_cost}
\end{lem} \vspace{-6pt}
\begin{pf}
	With the zero initial condition, zero input and $f(0,0)=0$, we have from \eqref{z_t} that $z(t)=0$ for all $t\in[0,\Delta]$. From the definition of the sampled-data cost \eqref{sd_cost} and positive definiteness of $\ell$, we obtain the statement. \qed
\end{pf}
\begin{thm}
	Suppose that $(x(0),r_0)\in\mathcal{XR}$ and that Assumptions \ref{ass_no_zeno}, \ref{ass_delta_tilde} and \ref{ass_min_max_choice} hold. Then, the closed-loop process state under application of Algorithm \ref{scheme_MPC} with the optimization problem $\mathcal{P}$ converges to $0$, i.e., $x(t)\rightarrow 0$ for $t\rightarrow\infty$. 
	\label{thm_tec_conv}
\end{thm} \vspace{-17pt}
\begin{pf}
	Consider again the feasible sequences \eqref{feas_inp} and \eqref{feas_state}. Consider the difference of value functions at $t_{k+1}$ and $t_k$
	\begin{align}
	&V^*(x(t_{k+1}),r_{k+1})-V^*(x(t_k),r_k) \nonumber \\
	&\le V(\tilde{\vec{x}}(t_{k+1}),\tilde{\vec{r}}(t_{k+1}),\tilde{\vec{u}}(t_{k+1}),\tilde{\vec{\Delta}}(t_{k+1}))-V^*(x(t_k),r_k) \nonumber \\
	&= \sum_{i=1}^{N-1} \lambda(x^*_i(t_k),u^*_i(t_k),\Delta^*_i(t_k)) + \underbrace{\lambda(0,0,\tilde{\Delta})}_{=0 \text{ (Lemma \ref{lem_sd_cost})}}  \nonumber \\ 
	&- \lambda(x(t_k),u^*_0(t_k),\Delta^*_0(t_k))- \sum_{i=1}^{N-1} \lambda(x^*_i(t_k),u^*_i(t_k),\Delta^*_i(t_k)) \nonumber \\
	&= - \lambda(x(t_k),u^*_0(t_k),\Delta^*_0(t_k)). \label{pf_tec_conv_1}
	\end{align}
	Using an induction on \eqref{pf_tec_conv_1} yields
	\begin{equation*}
	\begin{aligned}
	\lim_{k\rightarrow\infty} V^*(x(t_k),r_k) \le \hspace{2pt} &V^*(x(0),r_0) \\
	&- \sum_{k=0}^{\infty} \lambda(x(t_k),u^*_0(t_k),\Delta^*_0(t_k)).
	\end{aligned}
	\end{equation*}
	With $x(t)$ and $u(t)$, the closed-loop state and input trajectories resulting from an application of Algorithm \ref{scheme_MPC} with the optimization problem $\mathcal{P}$, we obtain further, using the definition of the sampled-data cost \eqref{sd_cost},
	\begin{align}
	\lim_{k\rightarrow\infty}V^*(x(t_k),r_k) \hspace{-1pt} &\le \hspace{-1pt} V^*(x(0),r_0) \hspace{-1pt} - \hspace{-1pt} \sum_{k=0}^{\infty} \int_{t_k}^{t_{k+1}} \hspace{-10pt} \ell(x(\tau),u(\tau)) \mathrm{d}\tau \nonumber \\
	=V^*(x(0),r_0)& - \lim_{k\rightarrow\infty} \int_{0}^{t_{k+1}} \ell(x(\tau),u(\tau)) \mathrm{d}\tau. \label{pf_tec_conv_2}
	\end{align}
	
	Recall that from Theorem \ref{lem_average_constraint}, it holds that $\lim_{k\rightarrow\infty} t_k = \infty$. Since $V^*(x(\infty),r_\infty)\ge 0$ due to the positive definite $\ell$, and since $V^*(x(0),r_0)$ is finite, we have from \eqref{pf_tec_conv_2} that
	\begin{equation*}
	\int_{0}^{\infty} \ell(x(\tau),u(\tau)) \mathrm{d}\tau
	\end{equation*}
	is also finite. We have that $\ell$ is continuous and positive definite, and $x(\cdot)$ is absolutely continuous. Further, $x(\cdot)$ and $\dot{x}(\cdot)$ are bounded from Assumption \ref{ass_standing}. Using Barbalat's Lemma (\cite{Khalil02}), we conclude that $x(t)\rightarrow 0$ as $t\rightarrow\infty$. \qed
\end{pf}

As demonstrated in this section, the resource dynamics and chosen minimum and maximum inter-sampling periods play a crucial role for recursive feasibility and convergence. To be precise, already for recursive feasibility, it is required by Assumption \ref{ass_delta_tilde} that the resource cost is such that after a transmission, the resource level may fill up again to the level before the transmission occurred. Furthermore, with Assumption \ref{ass_min_max_choice}, it is ensured that the sampling interval required to recover the resource is admissible in the MPC optimization problem. To establish convergence, one also needs to make sure that the sampling time instants are unbounded. In view of Assumption \ref{ass_no_zeno}, this can be ensured either by a larger-than-zero resource cost when choosing a zero inter-sampling period, or by a larger-than-zero minimum inter-sampling period.
\begin{remark}
	For the proposed resource-aware self-triggered MPC with resource constraints, we do not comment on stability in a continuous-time sense on top of convergence. To establish stability in continuous-time MPC, one usually verifies that the value function satisfies the relation $V^*(x(t_k+\delta))-V^*(t_k)\le 0$ for all $\delta\in[0,\Delta]$, where $\Delta$ denotes the sampling interval (cf. \cite{Chen98}). Subsequently, the value function is used as a Lyapunov function. For the proposed setup, this procedure is not possible. Due to the resource constraint $r\ge 0$, it cannot be guaranteed that $\mathcal{P}(x(t_k+\delta),\min\{r_k+p\delta-\mu(\delta),\overline{r}\}), \;\delta\in[0,\Delta^*_0(t_k))$ admits a solution, even though $\mathcal{P}$ was feasible at $t_k$. It can only be guaranteed that $\mathcal{P}(x(t_k+\Delta^*_0(t_k)),r_{k+1})$ is feasible, as demonstrated in the proof of Theorem \ref{thm_tec_rec_feas}.
	\label{rem_stability}
\end{remark}

\section{Numerical Example} \label{sec_num_ex}

For numerical simulation, we consider a double integrator
\begin{equation*}
\dot{x}(t) = \begin{bmatrix}0 &1 \\ 0 &0\end{bmatrix} x(t)  + \begin{bmatrix} 0 \\ 1\end{bmatrix}u(t), \; x(0) = \begin{bmatrix} 1 \\ 0 \end{bmatrix}
\end{equation*}
with the input constraint $u(t)\in[-2,2]$. The MPC optimization problem $\mathcal{P}$ with terminal equality constraint and $N=20$ is used here. The process cost penalizes quadratically the state and input
\begin{equation}
\ell(x,u)=x^\top \text{diag} (100,100)  x + u^2. \label{ex_cost}
\end{equation}

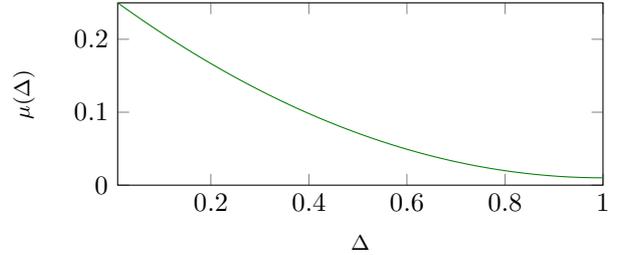
\begin{figure}
	\centering
%
%
\definecolor{mycolor1}{rgb}{0.00000,0.44700,0.74100}%
\begin{tikzpicture}

\begin{axis}[%
width=0.9\columnwidth,
height=4cm,
at={(0.758in,0.481in)},
xmin=0.01,
xmax=1,
ymin=0,
ymax=0.25,
axis background/.style={fill=white},
xlabel style={font=\color{white!15!black}, font=\small},
ylabel style={font=\color{white!15!black}, font=\small},
ylabel = {$\mu(\Delta)$},
xlabel = {$\Delta$}
]
\addplot [color=istgreen, forget plot]
  table[row sep=crcr]{%
0.01	0.25\\
0.02	0.24517649\\
0.03	0.24040196\\
0.04	0.23567641\\
0.05	0.23099984\\
0.06	0.22637225\\
0.07	0.22179364\\
0.08	0.21726401\\
0.09	0.21278336\\
0.1	0.20835169\\
0.11	0.203969\\
0.12	0.19963529\\
0.13	0.19535056\\
0.14	0.19111481\\
0.15	0.18692804\\
0.16	0.18279025\\
0.17	0.17870144\\
0.18	0.17466161\\
0.19	0.17067076\\
0.2	0.16672889\\
0.21	0.162836\\
0.22	0.15899209\\
0.23	0.15519716\\
0.24	0.15145121\\
0.25	0.14775424\\
0.26	0.14410625\\
0.27	0.14050724\\
0.28	0.13695721\\
0.29	0.13345616\\
0.3	0.13000409\\
0.31	0.126601\\
0.32	0.12324689\\
0.33	0.11994176\\
0.34	0.11668561\\
0.35	0.11347844\\
0.36	0.11032025\\
0.37	0.10721104\\
0.38	0.10415081\\
0.39	0.10113956\\
0.4	0.09817729\\
0.41	0.095264\\
0.42	0.09239969\\
0.43	0.08958436\\
0.44	0.08681801\\
0.45	0.08410064\\
0.46	0.08143225\\
0.47	0.07881284\\
0.48	0.07624241\\
0.49	0.07372096\\
0.5	0.07124849\\
0.51	0.068825\\
0.52	0.06645049\\
0.53	0.06412496\\
0.54	0.06184841\\
0.55	0.05962084\\
0.56	0.05744225\\
0.57	0.05531264\\
0.58	0.05323201\\
0.59	0.05120036\\
0.6	0.04921769\\
0.61	0.047284\\
0.62	0.04539929\\
0.63	0.04356356\\
0.64	0.04177681\\
0.65	0.04003904\\
0.66	0.03835025\\
0.67	0.03671044\\
0.68	0.03511961\\
0.69	0.03357776\\
0.7	0.03208489\\
0.71	0.030641\\
0.72	0.02924609\\
0.73	0.02790016\\
0.74	0.02660321\\
0.75	0.02535524\\
0.76	0.02415625\\
0.77	0.02300624\\
0.78	0.02190521\\
0.79	0.02085316\\
0.8	0.01985009\\
0.81	0.018896\\
0.82	0.01799089\\
0.83	0.01713476\\
0.84	0.01632761\\
0.85	0.01556944\\
0.86	0.01486025\\
0.87	0.01420004\\
0.88	0.01358881\\
0.89	0.01302656\\
0.9	0.01251329\\
0.91	0.012049\\
0.92	0.01163369\\
0.93	0.01126736\\
0.94	0.01095001\\
0.95	0.01068164\\
0.96	0.01046225\\
0.97	0.01029184\\
0.98	0.01017041\\
0.99	0.01009796\\
1	0.01007449\\
};
\end{axis}
\end{tikzpicture}%
 \vspace{-8pt}
	\caption{Energy cost function $\mu$ in the numerical example.}
	\label{fig_mu}
\end{figure}

The considered resource pertains to Example 2 in Section \ref{sec_average_constraints}, where the resource cost $\mu$ is related to the energy expended using a certain sampling interval $\Delta$. The minimum and maximum inter-sampling periods allowed by the processing unit are $\underline{\Delta}=\SI{0.01}{\second}$ and $\overline{\Delta}=\SI{1}{\second}$. Within these bounds, the energy cost is described in this example by
\begin{equation*}
\mu(\Delta) = 0.2449 (\Delta-\underline{\Delta})^2 - 0.4848 (\Delta-\underline{\Delta}) + 0.25.
\end{equation*}
A plot of this function can be found in Figure \ref{fig_mu}, where it can be seen that lowering the sampling time leads to a quadratically increasing energy cost. The resource allows a maximum level of $\overline{r}=0.5$ and is charged at a rate of $p=0.5$. With these parameters, $p\Delta-\mu(\Delta)=0$ is fulfilled for $\Delta\approx 0.176$, such that Assumptions \ref{ass_no_zeno}, \ref{ass_delta_tilde} and \ref{ass_min_max_choice} hold. The energy resource is full at the initial time, i.e., $r_0 = 0.5$.
\begin{figure}
	\centering
%
%
\begin{tikzpicture}

\begin{axis}[%
width=\columnwidth,
height=5cm,
at={(2.597in,1.085in)},
xmin=0,
xmax=5,
ymin=-2,
ymax=2,
axis background/.style={fill=white},
xlabel style={font=\color{white!15!black}, font=\small},
xlabel={Time $t$},
legend style={legend cell align=left, align=left, draw=white!15!black,legend columns=-1,font=\small,at={(0.97,0.97)}, anchor=north east,},
]

\addplot [color=istorange, mark=x, mark options={solid, istorange}]
table[row sep=crcr]{%
	0	1\\
	0.171139858347987	0.970711174628828\\
	0.342174066748529	0.882917130994297\\
	0.411876135516011	0.833299500124209\\
	0.54507975404547	0.736335644955273\\
	0.842130728685006	0.548409644488149\\
	1.10197333859084	0.422302430754291\\
	1.37796753446292	0.319987421731422\\
	1.65677872140024	0.24178372697285\\
	1.93758755911065	0.182326730667708\\
	2.2214750148698	0.137065985957048\\
	2.51012105669492	0.102548862292503\\
	2.80491892134553	0.0762511876195957\\
	3.10742630742308	0.0562588721892973\\
	3.44705051482648	0.0399711662728993\\
	3.77797305663372	0.0286584979839439\\
	4.09435363973729	0.020851035141147\\
	4.39732595266759	0.0153768308718678\\
	4.6900978173992	0.011456782509486\\
	4.97601511410859	0.00859511612673882\\
	5.25770811251069	0.00647567232004559\\
	5.5946186072236	0.117805683917021\\
	5.66883621332764	0.169804760546789\\
	5.81772423422803	0.277030787137319\\
	6.11257090874875	0.460703568739819\\
	6.37445137133578	0.585548965648325\\
	6.6507218586414	0.686047737035503\\
	6.92964555489947	0.762803190413524\\
	7.21062022174416	0.821162023656816\\
	7.49476842600006	0.865591953245191\\
	7.78377994860522	0.899476662799639\\
	8.07905102313903	0.925290463072743\\
	8.38204952835836	0.944905748697496\\
	8.72151643751333	0.960849791658418\\
	9.0521444494199	0.971921766116399\\
	9.36813098255377	0.979563037388263\\
	9.67072066720474	0.984922728006135\\
	9.96319285657913	0.988763021907616\\
	10.2489060592891	0.99156805365323\\
};
\addlegendentry{$x_1(t)$}

\addplot [color=istgreen, mark=square, mark options={solid, istgreen}]
  table[row sep=crcr]{%
0	0\\
0.171139858347987	-0.342279415840314\\
0.342174066748529	-0.684346128318988\\
0.411876135516011	-0.73935998966173\\
0.54507975404547	-0.716514201135745\\
0.842130728685006	-0.548763589621667\\
1.10197333859084	-0.421879476036868\\
1.37796753446292	-0.319549224661556\\
1.65677872140024	-0.241430380336527\\
1.93758755911065	-0.182039171333755\\
2.2214750148698	-0.136824827708596\\
2.51012105669492	-0.102340923162551\\
2.80491892134553	-0.0760706450764185\\
3.10742630742308	-0.0561067254633765\\
3.44705051482648	-0.0398093226942383\\
3.77797305663372	-0.0285611680385751\\
4.09435363973729	-0.0207937118735142\\
4.39732595266759	-0.0153429516805487\\
4.6900978173992	-0.0114359081390272\\
4.97601511410859	-0.00858153337767711\\
5.25770811251069	-0.00646636499903589\\
5.5946186072236	0.667354128034394\\
5.66883621332764	0.733905744370338\\
5.81772423422803	0.706452263508156\\
6.11257090874875	0.539434480176945\\
6.37445137133578	0.414018829590449\\
6.6507218586414	0.313520129545638\\
6.92964555489947	0.236848694672179\\
7.21062022174416	0.178554116539631\\
7.49476842600006	0.134169517978663\\
7.78377994860522	0.100317392778456\\
8.07905102313903	0.0745307552418713\\
8.38204952835836	0.054943715997213\\
8.72151643751333	0.0389920552425266\\
9.0521444494199	0.0279833615960841\\
9.36813098255377	0.020381175943411\\
9.67072066720474	0.015044292205342\\
9.96319285657913	0.0112166244971648\\
10.2489060592891	0.00841867215001843\\
};
\addlegendentry{$x_2(t)$}

\addplot[const plot, color=istblue, mark=o, mark options={solid, istblue}] table[row sep=crcr] {%
0	-1.99999824204798\\
0.171139858347987	-1.99999003519573\\
0.342174066748529	-0.789271571354109\\
0.411876135516011	0.171510269602266\\
0.54507975404547	0.564719949892907\\
0.842130728685006	0.488311419096277\\
1.10197333859084	0.370769577425257\\
1.37796753446292	0.280185473126625\\
1.65677872140024	0.211500497943797\\
1.93758755911065	0.159268550645364\\
2.2214750148698	0.11946778943512\\
2.51012105669492	0.0891128506553727\\
2.80491892134553	0.0659948170915877\\
3.10742630742308	0.0479865757913437\\
3.44705051482648	0.0339902945089043\\
3.77797305663372	0.0245509888402915\\
4.09435363973729	0.0179909515171427\\
4.39732595266759	0.0133450102696963\\
4.6900978173992	0.00998321820400874\\
4.97601511410859	0.00750877157273857\\
5.25770811251069	1.99999852663423\\
5.5946186072236	0.896709282736064\\
5.66883621332764	-0.184390125519556\\
5.81772423422803	-0.566456391623552\\
6.11257090874875	-0.478904189138667\\
6.37445137133578	-0.363769221334282\\
6.6507218586414	-0.274883188133719\\
6.92964555489947	-0.207472719114465\\
7.21062022174416	-0.156202284217132\\
7.49476842600006	-0.117130711243146\\
7.78377994860522	-0.0873320814688606\\
8.07905102313903	-0.0646440127830982\\
8.38204952835836	-0.0469903260803668\\
8.72151643751333	-0.033296312623244\\
9.0521444494199	-0.0240585748299985\\
9.36813098255377	-0.0176373617766415\\
9.67072066720474	-0.0130872877738042\\
9.96319285657913	-0.00979287033503337\\
};
\addlegendentry{$u(t)$}

\end{axis}
\end{tikzpicture}%
 \vspace{-8pt}
	\caption{Process state and input trajectories.}
	\label{fig_state}
\end{figure}
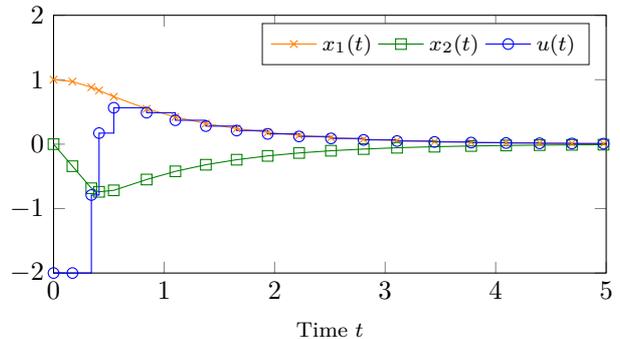
\begin{figure}
	\centering
%
%
\definecolor{mycolor1}{rgb}{0.00000,0.44700,0.74100}%
\definecolor{mycolor2}{rgb}{0.85000,0.32500,0.09800}%
\begin{tikzpicture}

\begin{axis}[%
width=\columnwidth,
height=5cm,
at={(0.758in,0.481in)},
xmin=0,
xmax=5,
ymin=0,
ymax=0.5,
ytick = {0,0.1,0.2,0.3,0.4,0.5},
yticklabels = {0,0.1,0.2,0.3,0.4,0.5},
axis background/.style={fill=white},
xlabel style={font=\color{white!15!black}, font=\small},
xlabel={Sampling Instants $t_k$},
ymajorgrids,
legend style={legend cell align=left,legend columns=-1, align=left, draw=white!15!black, font=\small, at={(0.97,0.97)}, anchor=north east}
]
\addplot [color=istgreen, draw=none, mark=x, mark options={solid, istgreen}]
  table[row sep=crcr]{%
0	0.171139858347987\\
0.171139858347987	0.171034208400542\\
0.342174066748529	0.0697020687674818\\
0.411876135516011	0.133203618529459\\
0.54507975404547	0.297050974639536\\
0.842130728685006	0.259842609905837\\
1.10197333859084	0.275994195872073\\
1.37796753446292	0.27881118693732\\
1.65677872140024	0.280808837710417\\
1.93758755911065	0.283887455759148\\
2.2214750148698	0.288646041825121\\
2.51012105669492	0.294797864650612\\
2.80491892134553	0.302507386077547\\
3.10742630742308	0.339624207403397\\
3.44705051482648	0.330922541807238\\
3.77797305663372	0.316380583103578\\
4.09435363973729	0.302972312930296\\
4.39732595266759	0.292771864731611\\
4.6900978173992	0.285917296709385\\
4.97601511410859	0.281692998402105\\
5.25770811251069	0.33691049471291\\
5.5946186072236	0.0742176061040434\\
5.66883621332764	0.148888020900391\\
5.81772423422803	0.29484667452072\\
6.11257090874875	0.261880462587021\\
6.37445137133578	0.276270487305627\\
6.6507218586414	0.278923696258071\\
6.92964555489947	0.280974666844686\\
7.21062022174416	0.284148204255899\\
7.49476842600006	0.289011522605159\\
7.78377994860522	0.295271074533809\\
8.07905102313903	0.302998505219335\\
8.38204952835836	0.339466909154972\\
8.72151643751333	0.330628011906562\\
9.0521444494199	0.315986533133872\\
9.36813098255377	0.30258968465097\\
9.67072066720474	0.292472189374392\\
9.96319285657913	0.285713202710019\\
};
\addlegendentry{$\Delta_k$}

\addplot [color=istorange, draw=none, mark=square, mark options={solid, istorange}]
  table[row sep=crcr]{%
0	0.178229968636522\\
0.171139858347987	0.178272857947258\\
0.342174066748529	0.221926352214387\\
0.411876135516011	0.193981871302216\\
0.54507975404547	0.131000877197484\\
0.842130728685006	0.144149485671504\\
1.10197333859084	0.138358593060527\\
1.37796753446292	0.137361690624244\\
1.65677872140024	0.136657098969193\\
1.93758755911065	0.135575066059163\\
2.2214750148698	0.13391171198312\\
2.51012105669492	0.131777790438038\\
2.80491892134553	0.129129708350505\\
3.10742630742308	0.1167881695617\\
3.44705051482648	0.119620972160805\\
3.77797305663372	0.124437834626949\\
4.09435363973729	0.128970944975021\\
4.39732595266759	0.132478514922609\\
4.6900978173992	0.134864182954107\\
4.97601511410859	0.136345870473109\\
5.25770811251069	0.11766763196128\\
5.5946186072236	0.219874022858336\\
5.66883621332764	0.187383924326135\\
5.81772423422803	0.131760933562164\\
6.11257090874875	0.143411803247199\\
6.37445137133578	0.138260644639078\\
6.6507218586414	0.137321955511019\\
6.92964555489947	0.136598697207006\\
7.21062022174416	0.135483634764629\\
7.49476842600006	0.133784417505536\\
7.78377994860522	0.131614412828585\\
8.07905102313903	0.128962003973427\\
8.38204952835836	0.116839048448433\\
8.72151643751333	0.119717504363431\\
9.0521444494199	0.124569800515267\\
9.36813098255377	0.129101597295435\\
9.67072066720474	0.132582333102826\\
9.96319285657913	0.134935568725504\\
};
\addlegendentry{$\mu(\Delta_k)$}
\addplot[draw=none, color=istblue, mark=o, mark options={solid, istblue}] table[row sep=crcr] {%
	0	0.5\\
	0.171139858347987	0.407339960537472\\
	0.342174066748529	0.314584206790485\\
	0.411876135516011	0.127508888959839\\
	0.54507975404547	0.000128826922352721\\
	0.842130728685006	0.0176534370446364\\
	1.10197333859084	0.00342525632605092\\
	1.37796753446292	0.00306376120156091\\
	1.65677872140024	0.00510766404597685\\
	1.93758755911065	0.00885498393199202\\
	2.2214750148698	0.0152236457524029\\
	2.51012105669492	0.0256349546818427\\
	2.80491892134553	0.0412560965691104\\
	3.10742630742308	0.0633800812573786\\
	3.44705051482648	0.116404015397378\\
	3.77797305663372	0.162244314140192\\
	4.09435363973729	0.195996771065031\\
	4.39732595266759	0.218511982555159\\
	4.6900978173992	0.232419399998356\\
	4.97601511410859	0.240513865398941\\
	5.25770811251069	0.245014494126885\\
	5.5946186072236	0.29580210952206\\
	5.66883621332764	0.113036889715746\\
	5.81772423422803	9.69758398065013e-05\\
	6.11257090874875	0.0157593795380024\\
	6.37445137133578	0.00328780758431366\\
	6.6507218586414	0.00316240659804959\\
	6.92964555489947	0.00530229921606595\\
	7.21062022174416	0.0091909354314032\\
	7.49476842600006	0.0157814027947236\\
	7.78377994860522	0.0265027465917672\\
	8.07905102313903	0.0425238710300874\\
	8.38204952835836	0.0650611196663281\\
	8.72151643751333	0.117955525795381\\
	9.0521444494199	0.163552027385232\\
	9.36813098255377	0.196975493436901\\
	9.67072066720474	0.219168738466951\\
	9.96319285657913	0.23282250005132\\
	10.2489060592891	0.240743532680826\\
};
\addlegendentry{$r_k$}
\end{axis}
\end{tikzpicture}%
 \vspace{-8pt}
	\caption{Chosen sampling intervals and resulting costs.}
	\label{fig_sampling_intervals_costs}
\end{figure}
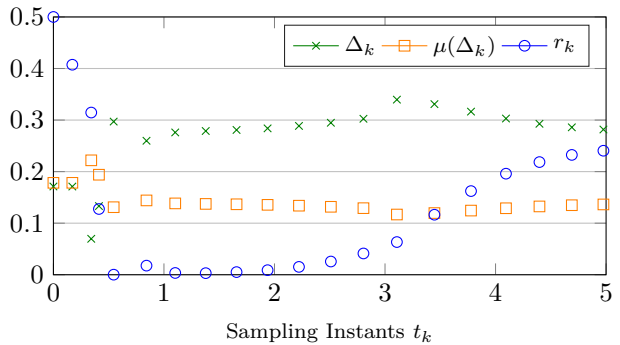

In Figure \ref{fig_state}, one can see the resulting process state and input trajectories. The states and the input converge to the origin. After initial feasibility, recursive feasibility and convergence are guaranteed by Theorems \ref{thm_tec_rec_feas} and \ref{thm_tec_conv}.

Figure \ref{fig_sampling_intervals_costs} shows the sampling intervals chosen by the control scheme and the resulting resource costs and resource levels at the sampling instants. To compensate the initial condition, one can see that the controller favors shorter intervals over a couple of sampling instants, at the expense of higher resource costs. By this measure, a high performance with respect to the process cost \eqref{ex_cost} can be achieved. However, it is only possible to use these very short inter-sampling periods until the stored resources are used up. After approximately \SI{0.5}{\second}, longer intervals must be used to comply with the resource constraint \eqref{resource_constraint}, although the plant is still far from the set point. After \SI{2.5}{\second}, the plant has almost converged and longer sampling intervals are used afterwards, such that the resource level rises again. Note, however, that this behavior cannot be expected in general, since it is not the goal in this work to penalize shorter inter-sampling periods as for example in \cite{Henriksson12,Li17}. Depending on $\mu$, it might even incur a lower cost to use short sampling intervals. Instead, the proposed scheme chooses a sampling interval that is favorable for the process cost $\ell$ and that complies with the resource constraint \eqref{resource_constraint}.

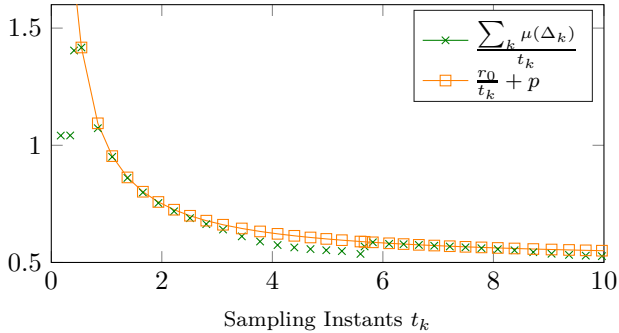
\begin{figure}
	\centering
%
%
\definecolor{mycolor1}{rgb}{0.00000,0.44700,0.74100}%
\definecolor{mycolor2}{rgb}{0.85000,0.32500,0.09800}%
\begin{tikzpicture}

\begin{axis}[%
width=\columnwidth,
height=5cm,
at={(2.597in,1.085in)},
xmin=0,
xmax=10,
xlabel style={font=\color{white!15!black}, font=\small},
xlabel={Sampling Instants $t_k$},
ymin=0.5,
ymax=1.6,
ylabel style={font=\color{white!15!black}},
axis background/.style={fill=white},
legend style={legend cell align=left, align=left, draw=white!15!black, font=\small}
]
\addplot [color=istgreen, draw=none, mark=x, mark options={solid, istgreen}]
  table[row sep=crcr]{%
0.171139858347987	1.04142874931051\\
0.342174066748529	1.04187564525684\\
0.411876135516011	1.40437653197239\\
0.54507975404547	1.41706061244746\\
0.842130728685006	1.07276922278867\\
1.10197333859084	0.950623192307854\\
1.37796753446292	0.860629859826217\\
1.65677872140024	0.798707563998505\\
1.93758755911065	0.753482746500211\\
2.2214750148698	0.718222735345961\\
2.51012105669492	0.688980943390337\\
2.80491892134553	0.663549791025984\\
3.10742630742308	0.640508535214372\\
3.44705051482648	0.611282379806358\\
3.77797305663372	0.589401295561583\\
4.09435363973729	0.574249382365143\\
4.39732595266759	0.564013452829004\\
4.6900978173992	0.557052242921028\\
4.97601511410859	0.55214737669613\\
5.25770811251069	0.548497463232388\\
5.5946186072236	0.536498983186143\\
5.66883621332764	0.568261473029418\\
5.81772423422803	0.585927590245523\\
6.11257090874875	0.579220450394926\\
6.37445137133578	0.577922343976027\\
6.6507218586414	0.574704310894674\\
6.92964555489947	0.571388600883658\\
7.21062022174416	0.568067523940385\\
7.49476842600006	0.564607546181877\\
7.78377994860522	0.56083127433389\\
8.07905102313903	0.556624983263463\\
8.38204952835836	0.551889323591107\\
8.72151643751333	0.543804821895577\\
9.0521444494199	0.537167764444626\\
9.36813098255377	0.532346314022233\\
9.67072066720474	0.529039331317407\\
9.96319285657913	0.526816453700608\\
10.2489060592891	0.5252960136252\\
};
\addlegendentry{$\tfrac{\sum_k \mu(\Delta_k)}{t_k}$}

\addplot [color=istorange, mark=square, mark options={solid}]
  table[row sep=crcr]{%
0.171139858347987	3.42158708571165\\
0.342174066748529	1.96124457867656\\
0.411876135516011	1.71395719947111\\
0.54507975404547	1.4172969575353\\
0.842130728685006	1.09373204535685\\
1.10197333859084	0.953731485590549\\
1.37796753446292	0.862853251252311\\
1.65677872140024	0.801790452485666\\
1.93758755911065	0.758052854256299\\
2.2214750148698	0.725075680191391\\
2.51012105669492	0.699193580192642\\
2.80491892134553	0.67825827199317\\
3.10742630742308	0.660904861623135\\
3.44705051482648	0.645051544167803\\
3.77797305663372	0.632346100013089\\
4.09435363973729	0.622119397588744\\
4.39732595266759	0.613705466772751\\
4.6900978173992	0.606607584631841\\
4.97601511410859	0.600482009908358\\
5.25770811251069	0.595098470531343\\
5.5946186072236	0.589371597083386\\
5.66883621332764	0.588201525178039\\
5.81772423422803	0.585944259278963\\
6.11257090874875	0.58179864208763\\
6.37445137133578	0.57843812288668\\
6.6507218586414	0.575179809143626\\
6.92964555489947	0.572153762560985\\
7.21062022174416	0.569342162618995\\
7.49476842600006	0.566713202007076\\
7.78377994860522	0.564236142761152\\
8.07905102313903	0.561888456771465\\
8.38204952835836	0.559651281981619\\
8.72151643751333	0.557329479750721\\
9.0521444494199	0.555235530408714\\
9.36813098255377	0.553372439062941\\
9.67072066720474	0.55170245498824\\
9.96319285657913	0.550184715602472\\
10.2489060592891	0.548785694503154\\
};
\addlegendentry{$\tfrac{r_0}{t_k}+p$}
\end{axis}
\end{tikzpicture}%
 \vspace{-8pt}
	\caption{Average resource cost and theoretical limit.}
	\label{fig_average_constraint}
\end{figure}
A set point change from $[0,0]^\top$ to $[1,0]^\top$ at \SI{5}{\second} was additionally implemented to demonstrate that the bound given by \eqref{no_zeno} holds whenever the MPC optimization problem is recursively feasible, independent of set point changes or other unforeseen operating conditions. In Figure \ref{fig_average_constraint}, the cumulated average resource cost from an application of the MPC is depicted, together with its theoretical limit as given by Lemma \ref{lem_average_constraint_finite}. We observe that especially right after the initial time and the set point change, the theoretical bound is very tight. After the plant has converged, the difference becomes larger. This complies very much with the discussion in Section \ref{sec_average_constraints}: because less resources were spent at other times (when the plant is close to the set point), it is possible to spend more when desired (right after a set point change). In other words, using less resources at times enables a short-time excess of resource usage, all while obeying transient constraints on the average usage.

\section{Conclusion} \label{sec_summary}

A method to incorporate awareness to a universal resource into MPC was presented in this paper. The considered scheme is self-triggered, i.e., it decides at each sampling instant at which point in time the controller should be activated next based on a joint optimization of the control inputs and future sampling instants. A novel point-in-time resource constraint was added to the MPC optimization problem, which determines whether a chosen sampling interval is feasible. The exact form of these constraints can be chosen by the user, such that the MPC scheme can be made aware toward an arbitrary resource. Interestingly, the point-in-time constraints imply that transient as well as asymptotic constraints on the average resource usage are also fulfilled. The exact form of the resource constraints also plays a role in establishing recursive feasibility of the MPC optimization problem and convergence of the process.

Future work could investigate a way to incorporate state constraints in closed loop despite the sampled-data formulation of the optimization problem. In a similar direction, also a way to handle disturbances on the process could be studied. Furthermore, one might imagine that the resource cost is also state-dependent, which would allow the user to incorporate more advanced resource constraints into the scheme.

{\small
\bibliography{bib_IFACWC20}   }

\end{document}